\documentclass[prl,twocolumn,showpacs,showkeys]{revtex4}
\usepackage{graphicx}
\usepackage{dcolumn}
\usepackage{bm}
\newcommand{\aanda}{Astron Astrophys}
\newcommand{\mnras}{Mon Not R Astron Soc}
\newcommand{\commentsastro}{Comments Astrophys Space Phys}
\newcommand{\araa}{Ann Rev Astron Astrophys}
\newcommand{\jgr}{J Geophys Res}
\begin{document}

\title{Particle acceleration in relativistic current sheets}
\author {J.G. Kirk}
\email{John.Kirk@mpi-hd.mpg.de}
\affiliation{Max-Planck-Institut f\"ur Kernphysik, Saupfercheckweg 1, 
D-69117 Heidelberg, Germany \\
\email{John,Kirk@mpi-hd.mpg.de}
}
\date{\today}
\begin{abstract}
Relativistic current sheets have been 
proposed as the sites of dissipation in pulsar winds,
jets in active galaxies 
and other Poynting flux dominated flows.
It is shown that the steady  versions of these structures 
differ from their nonrelativistic counterparts because
they do not permit transformation to a de~Hoffmann/Teller reference
frame, in which the electric field vanishes. 
Instead, their generic form is that of a true neutral sheet:
one in which the linking magnetic field component normal to the sheet 
is absent. 
Taken together with Alfv\'en's limit on the total 
cross-field potential, this suggests 
plasma is ejected from the sheet in the cross-field direction rather
than along it. 
The maximum energy to which such
structures can accelerate particles is derived, and used to compute the
maximum frequency of the subsequent synchrotron radiation. This can be
substantially in excess of standard estimates. 
In the magnetically driven gamma-ray burst scenario,
acceleration of electrons is possible to energies sufficient to enable
photon-photon pair production after an inverse Compton scattering event.
\end{abstract}
\maketitle

Energy release by the process of magnetic reconnection 
has been suggested as the mechanism 
responsible 
for the production of energetic particles in the
jets associated with active galactic nuclei (AGN) 
\cite{romanovalovelace92,leschbirk97,leschbirk98,
wiechenbirklesch98,litvinenko99,larrabeelovelaceromanova03} and
in pulsar winds \cite{michel71,lyubarsky96,lyubarskykirk01,
kirkskjaeraasengallant02,kirkskjaeraasen03}. It is thought that 
the same process may also
be at work in gamma-ray bursts 
\cite{thompson94,spruit99,spruitdaignedrenkhahn01,drenkhahn02,
drenkhahnspruit02}. 
Reconnection has a large
plasma physics, solar physics and magnetospheric physics orientated
literature, that is well-summarised in two recent monographs 
\cite{priestforbes00,biskamp00}. But the physical conditions
in the high energy astrophysics applications mentioned above 
are significantly different,
because relativistic effects are important.

In this letter, it is shown that the generic form of a stationary, relativistic
current sheet is that of a true neutral sheet. Typical boundary
conditions permit the sheet to be very extended along 
the magnetic field direction,
but a simple argument limits the extent in the direction perpendicular
to both the sheet normal and magnetic field vectors. This finding is used to
estimate the maximum possible energy to which particles can be
accelerated by the DC electric field of the sheet. The results are
applied to pulsar winds, gamma-ray bursts and jets from active
galactic nuclei, all of which may contain a magnetically powered 
relativistic outflow. It is shown that the commonly used estimate of
the maximum frequency of synchrotron radiation emitted by an
accelerated electron $h\nu\alt100\,$MeV can be substantially exceeded
when Poynting flux dominates the energy flow.

The current sheets at which reconnection and 
particle
acceleration takes place in astrophysics are relativistic in two senses:
Firstly, the magnetization parameter,
$\sigma$ (defined in Eq.~\ref{sigmadef}), 
is large and the Alfv\'en speed
$v_{\rm A}=c\sqrt{\sigma/(1+\sigma)}$ is close to $c$. Secondly,
the geometry of the current sheet at which magnetic energy is
dissipated is dictated by a highly relativistic plasma flow.
Particle acceleration depends crucially on both the magnetization parameter and
the field configuration. However, most analytic 
treatments utilize nonrelativistic models of current
sheets. Numerical simulations of current sheets with $\sigma\approx 1$
have been performed using a two-dimensional PIC code \cite{zenitanihoshino01}, 
but the field geometry was constrained to be close to the standard
Sweet-Parker configuration. Analytic work on relativistic reconnection 
\cite{blackmanfield94,lyutikovuzdensky02,kirkskjaeraasen03,lyutikov03} 
has not yet addressed the question of particle acceleration.

The relativistic effects associated with a large magnetization parameter
are readily appreciated. On the other hand, 
the geometrical effects of a relativistic flow are more subtle.
The situation is closely analogous to that of MHD shock
fronts, which can be classified into \lq\lq subluminal\rq\rq\ 
and \lq\lq superluminal\rq\rq\ according to whether the speed of the
intersection point of the magnetic field and the 
shock front is less or greater than $c$
\cite{drury83,begelmankirk90}. 
In each case, a Lorentz
transformation enables the shock to be viewed from a reference frame
in which it has a particularly simple
configuration.

Subluminal shocks permit a transformation to a de~Hoffmann/Teller
frame, where the electric field vanishes both upstream and
downstream. Nonrelativistic shocks, at which
the speed  $\beta_{\rm up}$
of the shock front in units of $c$ observed in the upstream
medium is small, $\beta_{\rm up}\ll 1$, are subluminal,
provided 
the angle between the magnetic field direction
and the shock normal in this frame $\theta_{\rm up}$ 
satisfies $\cos\theta_{\rm up}>\beta_{\rm up}$. 
Fine-tuning of the upstream magnetic field direction would be required
to violate this condition and, given that no MHD wave mode can race ahead
of the shock front, this seems unlikely. 

On the other hand, 
the simplest configuration for a superluminal shock is one in which
 the magnetic field lines are perpendicular to the shock normal both 
upstream and downstream --- an exactly \lq\lq perpendicular\rq\rq\
 shock. This 
is the usual configuration for shocks which are 
relativistic, i.e., those with
$\Gamma=\left(1-\beta_{\rm up}^2\right)^{-1/2}\gg1$. 
Then, fine-tuning of the upstream magnetic field direction such that 
$\sin\theta<1/\Gamma$ would be needed to {\em avoid} 
superluminal speed of the 
intersection point. 

Thus, the generic nonrelativistic shock can be pictured as a
stationary surface crossed by magnetic field lines along which 
the plasma flows. The generic relativistic shock, however, 
can be seen as a stationary surface through which a 
magnetic field orientated precisely perpendicular to the surface
normal is advected by the plasma.

\begin{figure}
\caption{\label{stripedsheet}(Color online)
The striped pattern of a pulsar wind. A magnetic dipole embedded in
the star at an oblique angle to the rotation axis introduces field
lines of both polarities into the equatorial plane. The current sheet 
separating these regions is shown. In the inset, 
an almost planar portion of this sheet (dashed line) is shown,
together with the magnetic  
field lines, assuming they undergo reconnection.}
\includegraphics[bb=0 0 759 560,width=7 cm]{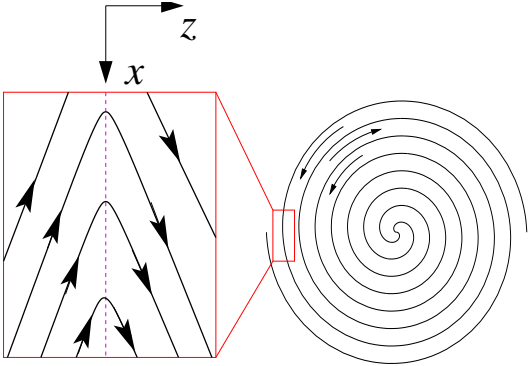}
\end{figure}

To understand the analogy with current sheets, it is best
to consider a specific example: 
the relativistic MHD wind driven by a pulsar --- a rotating
neutron star in
which a magnetic dipole is embedded with axis oblique to the rotation
axis. Approximating the dipole by a split-monopole, an asymptotic
solution is available, valid at large distance from the star
\cite{bogovalov99}: the wind pulls out the 
field lines into a striped pattern \cite{coroniti90}, of wavelength 
$2\pi\beta r_{\rm L}$,
($r_{\rm L}=c/\Omega$ is the light-cylinder radius, $\Omega$ the
rotation speed of the star) that propagates radially at a speed 
$\beta c$ close to $c$. 
Figure~\ref{stripedsheet} shows the equatorial plane where
a thin helical current sheet separates regions of magnetic field that
emerge from opposite magnetic poles. 
This is the most widely discussed scenario for a pulsar wind, although
no solution exists connecting it to realistic boundary conditions at
the stellar surface and alternative pictures involving currents which do
not collapse into thin sheets have also been proposed
\cite{kuijpers01}. 
In ideal MHD,
the current sheet is a neutral sheet in which the magnetic field
vanishes. The inset in Fig.~\ref{stripedsheet} shows that field line
reconnection implies a finite \lq\lq linking\rq\rq\ field component
perpendicular to the sheet. 
The term \lq\lq reconnection\rq\rq\ can be misleading here, since the
field line configuration is stationary in the corotating
frame. However, plasma certainly leaves these field lines, so that the
process qualifies as reconnection under a more general 
definition \cite{hesseschindler88,biskamp00}.

At a large distance $r\gg r_{\rm L}$ 
from the star, the
sheet can be considered as locally plane. Then the linking component
is in the $\bm{z}$ direction, and the reversing component in the
$\pm\bm{x}$ direction. The idealised steady reconnection picture
discussed here implies 
an electric field $E_y$ in the $\bm{y}$ direction
that causes the magnetized plasma to
drift into the current sheet from both positive and negative $z$.
A shearing component of the magnetic
field in the $\bm{y}$ direction can also be added, but this vanishes
in the equator if the magnetic and rotational axes are orthogonal and 
will be ignored here, in which case the electric field remains along the
$y$ axis.
 
In a stationary flow, 
reconnection balances the rate at
which magnetic flux is drawn out from the star, implying that 
the point at which a given magnetic field intersects the current sheet 
must corotate. However, far from the star, 
at $r\gg r_{\rm L}$, 
the corotation velocity is large compared to $c$ and 
the relativistic configuration of
Fig.~\ref{stripedsheet} implies a drift speed
in the sheet that is formally superluminal: $E_y/B_z\gg1$. 
This same situation also arises in
the case of extra-galactic jets, provided they have zero net poloidal flux 
i.e., they draw out magnetic flux from 
a central object\cite{romanovalovelace92}. 
In contrast, the current
sheets usually considered in laboratory, solar and magnetospheric applications
are assumed to be of finite length in the $\bm{x}$ direction and to 
have subluminal drift speeds at their edges.

A two-dimensional, stationary configuration has $E_y$ 
constant and, for simplicity, it is usual to
assume $B_z$ is constant too. Then it is possible to simplify the
fields by performing a Lorentz boost: a subluminal sheet can be 
transformed by a boost in the $\bm{x}$ direction at speed $cE_y/B_z$
into the \lq\lq de~Hoffmann/Teller\rq\rq\ frame in which the
electric field vanishes.  
In this frame the orbits are complex, some being trapped in the sheet
and others escaping along the magnetic field lines
after one or more passages through the plane $z=0$. This is the
standard configuration for the study of orbits in nonrelativistic
current sheets \cite{buechnerzelenyi89,chen92}.
The particle energy is a constant of motion, since $\bm{E}=0$, 
and no acceleration results 
unless an additional scattering process is invoked. It is then easy to
see that in any other reference 
frame the acceleration process can be regarded
as one of reflection off the magnetic field structure anchored in the
sheet. This implies that ions are accelerated more efficiently than
electrons, as observed in, for example, the geomagnetic tail
\cite{daglis94}.

Superluminal sheets, on the other hand, are
transformed by a boost along $\bm{x}$ with speed 
$cB_z/E_y$ into a configuration with
vanishing linking field $B_z$, but finite $E_y$. 
(This is, in fact, a true neutral sheet: the field configuration originally
used by \citet{speiser65} to discuss the orbits of nonrelativistic
particles.) 
In this case all trajectories are trapped in the sheet and
acceleration in the $\pm\bm{y}$ direction ensues, according to the
sign of the charge, as soon as a particle drifts into the acceleration
zone around $z=0$, where $\left|E_y/B_x\right|>1$.

Particle acceleration in current sheets with finite $B_z$ has been
extensively investigated \cite{syrovatskii81}, but 
the vanishing of $B_z$ has important implications. Most discussions of
reconnection treat a Sweet-Parker or Petschek configuration in which 
the length of the current sheet in the $x$ direction determines
the dissipation rate. The linking 
field $B_z$ is then crucial for the determination of the spectrum
of accelerated particles, 
and, especially, the
maximum permitted energy
\cite{litvinenko99,larrabeelovelaceromanova03}, since it is 
responsible for ejection of particles from the acceleration zone 
(see also \cite{zenitanihoshino01}).
However, relativistic current sheets, such as that illustrated in
Fig.~\ref{stripedsheet} can extend over large distances in the $x$
direction, depending on the nature of the boundary conditions, and 
the linking field plays no role in ejecting particles. Instead,
acceleration in a relativistic sheet is controlled by its
finite extent in the direction parallel to the 
electric field $E_y$. This is limited 
not by the boundary conditions, but by local parameter values,
as described by \citet{alfven68} and \citet{vasyliunas80}.

The maximum energy to which a particle can be accelerated by the DC
field in a
relativistic current sheet is determined by the product of the 
electric field $E_y$ and the maximum extent $\Delta y$
in the $\bm{y}$ direction
of the sheet. Alfv\'en showed that $E_y\times\Delta y$ must be finite 
by observing that charged particles 
which drift into the sheet through a surface of area 
$\Delta x\times\Delta y$ must leave it in opposite directions along 
the $y$ axis. The current $I_y$ in the $\bm{y}$ direction, 
part of which could be carried also 
by particles entering the sheet along this axis, must therefore exceed 
the rate of inflow through $\Delta x\times\Delta y$ times the
particle charge. Amp\'ere's law, which relates the strength of the
reversing component of the field to the current $I_y$ leads to 
the condition
$|q E_y|\Delta y< B_x^2/(4\pi n_\pm)$,
where $n_\pm$ is the 
number density of particles of charge $q$ outside
the sheet. This argument, originally limited to stationary sheet
configurations, can also 
be applied to evolving ones \cite{vasyliunas80}.

The magnetization parameter $\sigma$ can be defined in the local
plasma rest frame as the ratio of the magnetic enthalpy density 
to the particle enthalpy density (including rest-mass) $w$:
\begin{eqnarray}
\sigma&=&\frac{B^2}{4\pi w}.   
\label{sigmadef}
\end{eqnarray}
This quantity equals the ratio of the Poynting flux to the kinetic
energy flux of a cold plasma observed in a reference frame moving
perpendicular to the magnetic field direction.
Assuming the plasma consists of cold electrons and positrons, 
and that $\sigma\gg1$, the
maximum Lorentz factor $\gamma_{\rm max}$ after acceleration is,
following Alfv\'en,
\begin{eqnarray}
\gamma_{\rm max}&=&2\sigma,
\label{pairplasmalimit}
\end{eqnarray}
whereas a cold electron-proton plasma gives
\begin{eqnarray}
\gamma_{\rm max}\,\approx\,\sigma\ &&\textrm{ for protons}
\\
\gamma_{\rm max}\,\approx\,\sigma M/m\ &&\textrm{ for electrons},
\label{leschlimit}
\end{eqnarray}
with $M$ and $m$ the proton and electron masses, respectively.
It is interesting to note that in a plasma in which the magnetic field
and particle {\em rest mass} are in rough equipartition
($\sigma\approx1$), the upper limit given by Eq.~(\ref{leschlimit})
coincides with that quoted by \citet{leschbirk97}. However, this
situation arises only in relativistic plasmas. In the interstellar
medium, for example, $\sigma\approx10^{-9}$ or smaller, in which case 
the upper limit on the energy gain reduces to $M v_{\rm A}^2$. 
Standard estimates of the interstellar magnetic field and particle density 
($1\,\mu$G, $1$~proton/cm$^{3}$) imply that electrons can be
accelerated, at most, to
only mildly relativistic energies.
In this case, and in solar system applications, direct acceleration by
the DC field may be masked by particle acceleration in the turbulence
fed by reconnection or the associated shocks \cite{cargill01}.

In addition to the limit imposed by finite $\Delta y$, radiative loss
processes suffered during acceleration may also affect the maximum particle
(and especially electron) energy. In the absence of a
linking magnetic field component, the particle trajectories, as first
found in the nonrelativistic limit 
by Speiser \cite{speiser65}, are attracted to the 
acceleration zone around the $z=0$ plane,
where they undergo essentially one-dimensional acceleration in the
electric field. In this case the losses by synchrotron radiation
(or, more precisely, radiation caused by the
electromagnetic fields of the sheet)
are unimportant. For large $\sigma$, this enables synchrotron photons
of relatively large energy 
$h\nu_{\rm max}\approx \left(\hbar eB/mc\right)\gamma_{\rm max}^2$ 
to be generated when the accelerated
particles leave the current sheet after traversing the distance $\Delta
y$.
This can substantially exceed the frequently quoted (and
magnetic field independent)
 upper limit of 
$h\nu_{\rm max}\approx 100\,$MeV obtained by setting the synchrotron
cooling rate equal to the inverse of the gyro period. 
For example, a relativistic current sheet in 
the wind of the Crab pulsar just outside the light
cylinder ($B\approx 10^6\,$G, $\sigma\approx 10^4$ 
\cite{kirkskjaeraasen03}) is 
capable of accelerating electrons which 
subsequently emit synchrotron 
photons of energy 
up to $50\,$GeV, providing a plausible 
production mechanism \cite{lyubarsky96} for the flux of
GeV gamma-rays observed by the EGRET experiment \cite{nolanetal93}.

Unlike synchrotron radiation, inverse Compton scattering 
on an ambient radiation field is not suppressed for linearly
accelerated particles, and can limit the maximum energy, especially
near a luminous source such as an AGN. 
Parameterising the accelerating electric field by
$\eta=\left|E_y/B_x\right|<1$, this limit can be written
for a current sheet embedded in a jet that emerges from the AGN 
with bulk Lorentz factor
$\Gamma=\Gamma_{10}/10$ as
\begin{eqnarray}
\gamma_{\rm max}&\approx&
10^5\left(\eta L_{\rm P47}\right)^{1/4}
\left(\Gamma_{10}R_{12}/L_{\rm bb47}\right)^{1/2},
\label{agnestimate}
\end{eqnarray}
where $L_{\rm P47}$ and $L_{\rm bb47}$ are the luminosity carried by Poynting
flux in the jet and the luminosity of the central photon source,
respectively, in units of $10^{47}\textrm{erg\,s}^{-1}$, and
  $R=10^{12}R_{12}\,$cm is the distance of the sheet 
from this source. Equation~(\ref{agnestimate}) assumes that the
Compton scattering takes place in the Thomson limit, which is valid
for $\gamma_{\rm max}\alt 10^5\Gamma$. 

In the physical 
conditions thought prevalent at the photosphere of a
magnetically accelerated gamma-ray burst outflow
\cite{drenkhahnspruit02}
($L_{\rm P47}\approx 10^3$, $L_{\rm bb47}\approx50$ and
$\Gamma\approx100$ at the photosphere $R=10^{11}\,$cm), 
Eq.~(\ref{agnestimate}) indicates a maximum electron Lorentz factor 
(in the comoving frame) of $\gamma_{\rm max}\approx4\times10^5$. 
However, since the photon field in this
case has a relatively high temperature ($\approx1\,$keV), we conclude that 
Compton scattering is unable to limit the maximum electron 
energy to a value such that the Thomson limit applies, and,
consequently, that photon-photon pair creation is likely to ensue.
A feedback mechanism appears possible here, since pair creation will
depress the magnetization parameter $\sigma$, and, hence, the limit
given by Eq.~(\ref{pairplasmalimit}). 

Relativistic current sheets are known to be unstable to the growth
of the tearing mode \cite{zelenyikrasnoselskikh79} and other
instabilities are also likely to 
operate (see, for example, \cite{daughton99}). 
On scale lengths comparable to the sheet thickness it is likely that 
an oscillating component of $B_z$ will be generated. 
As discussed in
standard nonrelativistic pictures, 
this may, on small scales,
cause breakup of the sheet into a series of magnetic islands and 
X-points arranged alternately along the $x$-axis, 
providing stochastic deflections
of the particle trajectories within the sheet. Such a 
scattering mechanism is, in fact, 
essential for the formation of 
an equilibrium structure such as the Harris sheet
\cite{harris62}, and it may also play a role in enhancing the 
effect of synchrotron losses.
However, in a steady relativistic sheet
the boundary conditions do not permit this scattering process 
to evacuate the plasma from the dissipation region, because they prevent the
formation of a stationary pattern with outflow along the
$x$ axis. 
On the other hand, a quasi-stationary pattern containing 
relativistic sheets 
of finite extent in the $y$ direction does 
appear possible, and enables estimates of the maximum permitted energy
of particles accelerated by the DC field to be made. 
In reality, relativistic sheets may dissipate in a highly
non-stationary fashion as suggested in the nonrelativistic case by,
for example, 
solar observations
\cite{cargill01}, but clarification of this 
must await the results of 
numerical simulations of the generic relativistic sheet
configuration.
\begin{acknowledgments}
I thank Y.~Lyubarsky for helpful comments on the manuscript.
\end{acknowledgments}

\end{document}